\documentclass[%
 reprint,
superscriptaddress,
 amsmath,amssymb,
prl,
]{revtex4-2}

\usepackage{graphicx}
\usepackage{mathtools}
\usepackage{dcolumn}
\usepackage{bm}
\usepackage{physics}
\usepackage[hidelinks, colorlinks=false, pdfborder={0 0 0}]{hyperref}
\usepackage{empheq}

\newcommand{\cri}{CrI\textsubscript{3} {}}
\newcommand{\mzcri}{$M_z$-CrI\textsubscript{3} {}}
\newcommand{\mote}{MoTe\textsubscript{2} {}}
\newcommand{\mztmote}{$M_z$-$T_d$-MoTe\textsubscript{2} {}}
\newcommand\numberthis{\addtocounter{equation}{1}\tag{\theequation}}

\begin{document}

\title{
 Ultrafast switchable polar and magnetic orders by nonlinear light-matter interaction
}

\author{Haoyu Wei}
\affiliation{Department of Materials Science and Engineering, Massachusetts Institute of Technology, Cambridge, Massachusetts 02139, USA}%
\affiliation{Tsien Excellence in Engineering Program, Tsinghua University, Beijing 100084, China}
\affiliation{Department of Applied Physics, Yale University, New Haven, Connecticut 06520, USA}
\author{Daniel Kaplan}
\email{d.kaplan1@rutgers.edu}
\affiliation{Center for Materials Theory, Department of Physics and Astronomy, 
Rutgers University, Piscataway, NJ 08854, USA}%
\author{Haowei Xu}%
\affiliation{Department of Nuclear Science and Engineering, Massachusetts Institute of Technology, Cambridge, Massachusetts 02139, USA}%

\author{Ju Li}
\email{liju@mit.edu}
\affiliation{Department of Materials Science and Engineering, Massachusetts Institute of Technology, Cambridge, Massachusetts 02139, USA}%
\affiliation{Department of Nuclear Science and Engineering, Massachusetts Institute of Technology, Cambridge, Massachusetts 02139, USA}%

\date{\today}%

\begin{abstract}
An outstanding challenge in materials science and physics is the harnessing of light for switching charge order in e.g., ferroelectrics. Here we propose a mechanism through which electrons in ferroelectric bilayers excited with light cause \textit{ionic} structural transitions. Using perturbation theory within a many-body formalism, we show that the ionic coupling is mediated by a resonant change in electronic occupation functions, ultimately governed by the quantum geometric tensor (QGT) of the ground state. Furthermore, we show that such transitions are generally accompanied by multiferroic order switching. We demonstrate two examples of light-induced structural and polarization switching under this mechanism using first-principle calculations on bilayer \cri and \mote. We show that the two materials can switch between atomic stackings with a light intensity threshold of only 10\textsuperscript{1}-10\textsuperscript{2}GW/cm\textsuperscript{2}, a value 1-3 orders of magnitude lower than that required by direct light-ion coupling thanks to the superior efficiency of resonant light-electron coupling. Since such switching is fast, highly controllable, contactless, and reversible, it is promising for use in optically controlled nonvolatile memory, nanophotonics and electronic devices. 
\end{abstract}

\maketitle
\paragraph{Introduction.---} 
Switchable and reversible engineering of quantum materials with light is a long-sought goal of materials physics \cite{juraschek2017ultrafast,radaelli2018breaking,disa2021engineering}. A key ingredient for efficient switching is the existence of a manifold of states separated by small energy gaps that can be overcome by the energy pumped with light.
One common example of matter with competing orders with small gaps is the multiferroics \cite{spaldin_renaissance_2005}. Over past decades, multiferroics have elicited extensive research interest for potential applications, with the ultimate goal of converting them into practical devices. The possibility of using multiferroics -- and more specifically the diverse ferroic or magnetic orders found in them -- as  magnetically controllable ferroelectric memory, switches and sensor has been proposed \cite{spaldin_advances_2019} and is the subject of immense experimental effort. 

Recent work has explored the possibility that ferroic order can be used as a switch. For instance, ferroelastic transitions have been been proposed theoretically \cite{li_ferroelasticity_2016} and supported experimentally \cite{wang_formation_2019} in the 1T-to-1T' transition of monolayer transition metal dichalcogenides (TMD). Ferromagnetism has also been seen in a range of 2D materials including \cri \cite{huang_layer-dependent_2017}, $\mathrm{Cr}_2\mathrm{Ge}_2\mathrm{Te}_6$\cite{gong_discovery_2017}, CrXBr \cite{wang_electrically_2020,jiang_screening_2018,telford_layered_2020,lee_magnetic_2021}. Moreover, an inversely stacked \cri has been predicted and shown to also be ferroelectric, realizing a multiferroic state \cite{ji_general_2023,qian2024coexistence}. Beyond intrinsic properties of 2D stacks, experimentally observed sliding ferroelectricity \cite{yasuda_stacking-engineered_2021,vizner_stern_interfacial_2021} offers a way of realizing tunable out-of-plane ferroelectricity in bilayers that is switched by inter-layer shifts of the 2D constituents composing the structure. However, these works neglected contributions arising from quantum coherence and focused only on the classical dielectric response. A unification of the electronic and structural transitions is also missing. 

The use of light to tune between and induce multiferroic behavior locally, reversibly and in a switchable manner is particularly attractive. Previous works have considered non-soft, sliding transitions, such as optical tuning of AB stacked h-BN \cite{xu_optomechanical_2019} and SnO \cite{zhou_opto-mechanics_2018}. Light-induced magnetization reversal was separately considered in CrI\textsubscript{3} \cite{xu_light-induced_2021,zhang_all-optical_2022}, without the possbility of structural transiitons.

\begin{figure}[t]
\includegraphics[width=\linewidth, page=1]{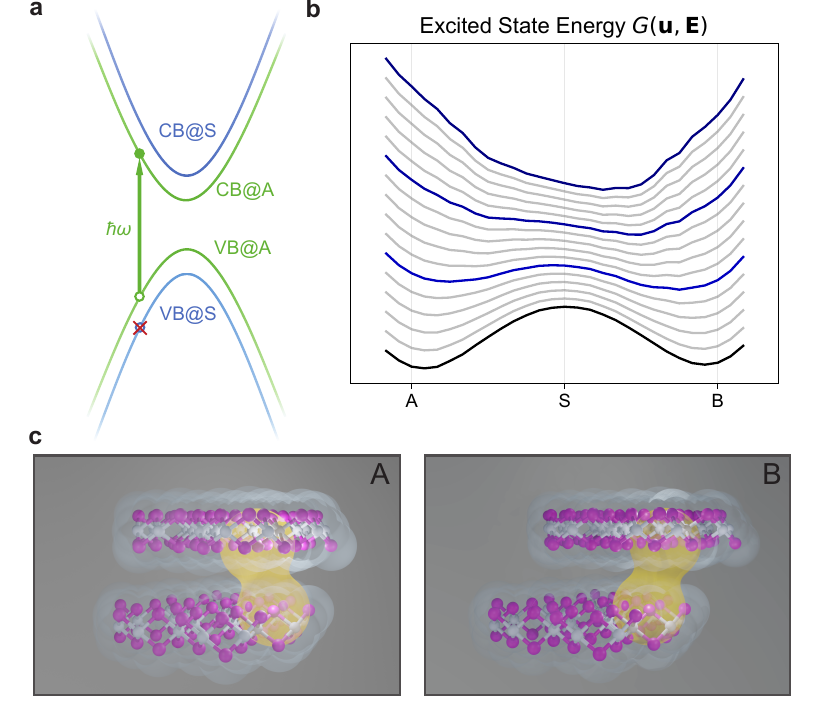}
\caption{\label{fig:1} Principles of Light-driven ferroic order switching. (a) Light generates excited states dependent on ionic order. CB, conduction band; VB, valence band. (b) The system's Born-Oppenheimer (BO) surface changes as the system is excited from the ground state (black curve in the bottom) to excited states (grey and blue curves). (c) Two stackings of bilayer \cri that can switch to each other from light pumping (excited electron cloud in yellow). In (a)(b)(c), A,B, labels for two locally stable ionic orders; S, label for the ionic order at saddle point in the switching path.}
\end{figure}

In this work we demonstrate that resonant visible light, inducing interband electronic transitions, can drive structural transitions via sliding. 
We concretely show how light induces a sliding structural transition and hence switches the polarization in bilayer CrI\textsubscript{3} and MoTe\textsubscript{2} -- and more broadly van der Waals (vdW) materials -- where two layers are related by a mirror symmetry $(M_z)$. These vdW heterostructures  \cite{huang2017layer,niu2019coexistence,an2020ferroic,yuan2019room,jindal2023coupled,qian2024coexistence,wang_interfacial_2022,bennett2024stacking,yasuda2024ultrafast} have been shown to exhibit robust ferroelectricity.

Schematically, our physical mechanism is described in Fig.~\ref{fig:1}(b). The free-energy in equilibrium is symmetric between two states (two atomic stackings, as shown in Fig.~\ref{fig:1}(c)-Fig.~\ref{fig:1}(d)). Upon illumination (curves of different color in Fig.~\ref{fig:1}(b)) the nearly-symmetric free energy deforms and softens into a single well, causing in instability in the atomic structure which leads to sliding. The control over which stacking softens depends on the matrix elements of the electronic states (and their associated quantum geometry), as shown in Fig.~\ref{fig:1}(a).

Employing first-principle calculations, we find that specifically chosen light frequency and polarization can independently and reversibly trigger multiferroic transition in these materials. To achieve this in \cri, we trigger its ferromagnetic order switching separately from ferroelastic order switching. We propose that ferroelastic and accompanying ferroelectric switching can be achieved by triggering interlayer sliding through the light-electron-ion coupling pathway using above-bandgap light, which we directly show by calculating the free energy from a many-body perspective. We show that ferromagnetic switching occurs through the nonlinear Edelstein effect, as introduced in Ref.~\cite{xu_light-induced_2021}. Here we demonstrate that the ferroelastic and ferromagnetic effects are governed by contrasting elements of quantum geometry. 

Our findings establish a new method of controlling crystal structure with optics, leading to ultrafast, reversible and widely applicable switching of electronic and ionic order in solids.

\paragraph{Theory of Nonlinear Electron-mediated Light-Ion Coupling.---}
We begin by introducing the theoretical framework of Nonlinear Electron-mediated Light-Ion Coupling (NELIC). Under an incident electric field $\mathbf{E}(t) = \mathbf{E}_0\cos{\omega t}$, the free energy $G$ of an electron-ion system is given by,
\begin{equation}
    G(\mathbf{u},\mathbf{E}) = \Phi(\mathbf{u},\mathbf{E}) - \mathbf{E} \cdot \mathbf{p}(\mathbf{u}) ,
    \label{eq:free-energy}
\end{equation}
where $\mathbf{p}$ is the dipole moment of the electron-ion system, and $\mathbf{u}$ denotes the positions of all ions $\mathbf{u}= \lbrace {\mathbf{u}_i} \rbrace$. Here $\Phi(\mathbf{u},\mathbf{E})$ is the Born-Oppenheimer (BO) energy surface for a given ionic configuration and electric field $\mathbf{E}$. The free energy $G(\mathbf{u},\mathbf{E})$ will thus be the effective potential felt by the ionic degrees of freedom, consistent with the Born-Oppenheimer approximation \cite{essen1977physics,worth2004beyond}.
While the free energy $G(\mathbf{u},\mathbf{E})$ generally contains many harmonics of $\omega$, only its static component will be relevant for significant ionic motion, as the rate of ionic motion frequency is typically $10^{-2}\sim 10^{-3}$ lower than optical frequencies $\omega$ at which electrons are resonant. The effect of the static component accumulates in time, resulting in ionic motion.

Working in the Kohn-Hohenberg formalism of DFT \cite{parr1994density,Argaman_2000,giustino2014materials}, we express Eq.~\eqref{eq:free-energy} using its density-related constituents
\begin{align*}
    &\Phi(\mathbf{u},\mathbf{E})
    = T_{e}[\rho(\mathbf{r},\mathbf{u},\mathbf{E})]
    + V_{e-e}[\rho(\mathbf{r},\mathbf{u},\mathbf{E})]
    ~ + \\ & \sum_i\int \frac{-Z_i e^2 \rho(\mathbf{r},\mathbf{u},\mathbf{E})}{|\mathbf{r}- \mathbf{u}_i|}  \textrm{d}\mathbf{r}
    + \sum_{i>j} \frac{Z_i Z_j e^2}{|\mathbf {u}_i - \mathbf {u}_j|}
    \numberthis
    \label{eq:interaction-energy}
\end{align*}
Here, $T_e(\rho)$ is the kinetic energy density functional, $V_{e-e}$ is the electron-electron interaction functional. $\rho(\mathbf{r},\mathbf{u},\mathbf{E})$ is the many-body electronic density as a function of local coordinate $\mathbf{r}$, and electric field $\mathbf{E}$. The summation $i,j$ refers to ionic position. The third term, $V_{e-i} = \sum_i\int \frac{-Z_i e^2 \rho(\mathbf{r},\mathbf{u},\mathbf{E})}{|\mathbf{r}- \mathbf{u}_i|}  \textrm{d}\mathbf{r}$ is the ion-electron coupling that feeds back the effect of light onto the ionic motion. We drop the fourth term as it does not depend on the electronic density and is unaffected by the application of light. 
Applying the Kohn-Sham approximation \cite{parr1994density},
$T_{e} + V_{e-e} + V_{e-i}$ can be evaluated using single-electron Hamiltonian in the Bloch basis $H|u_{n\mathbf{k}}\rangle = \hbar \omega_{n\mathbf{k}}|u_{n\mathbf{k}}\rangle$ in the Bloch basis of the solid,
\begin{align*}
    &T_e + V_{e-e} + V_{e-i}
    = \textrm{Tr}({\rho  H }) 
    = \sum_n  \int_\mathbf{k}\rho_{n\mathbf{k}} \hbar \omega_{n \mathbf{k}} \numberthis
    \label{eq:band-theory}
\end{align*}
where $\rho_{n\mathbf{k}}$ is the density matrix of the $n$-th Bloch band at momentum $\mathbf{k}$. We use the notation $\int_{\mathbf{k}} = \frac{1}{(2 \pi)^d}\int \textrm{d}^d \mathbf{k}$ where $d$ is the spatial dimension.

As we detail in the Supplementary Information (SI), by expanding $\Phi(\mathbf{u},\mathbf{E})$ to leading order in $\mathbf{E}$, the unperturbed BO energy surface in equilibrium $\Phi(\mathbf{u},0)$ is affected at finite $\mathbf{E}$ purely by a change in \textit{electronic} occupation. We find that,
\begin{align}
    \Phi(\mathbf{u},\mathbf{E})
    = \Phi(\mathbf{u},0)
    + \sum_n  \int_\mathbf{k} \delta\rho_{n\mathbf{k}}(\mathbf{E}) \hbar{\omega}_{n\mathbf{k}}
    \label{eq:band}
\end{align}
and,
\begin{align*}
 \delta &\rho_{n\mathbf{k}}(\mathbf{E})
    = \\ &  \frac{i e^2V_\mathrm{u.c.}}{\hbar^2 \omega^2} \sum_{lab} 
        \frac{1}{\gamma_{ln}}\frac{f_{ln} v_{nl}^a v_{ln}^b }{\omega_{ln} - \omega + i\gamma_{ln}}
     E_a(\omega) E_b(-\omega) + \mathrm{c.c.}
    \numberthis
    \label{eq:density_matrix}
\end{align*}
Eq.~\eqref{eq:density_matrix} represents the first non-vanishing term in the perturbation theory of static change in Bloch state occupation number due to light illumination. $\gamma_{ln}$ is the electronic decay rate from level $l$ to $n$, $\omega$ is the frequency of light. Here, $v^a_{ln} = \langle u_{l\mathbf{k}}|\hat{v}^a|u_{n\mathbf{k}}\rangle$ is the velocity operator in the $a(=x,y,z)$ direction evaluated under Bloch basis, and we use the shorthands $f_{ln} = f_{l\mathbf{k}} - f_{n\mathbf{k}}$ for the equilibrium occupation number difference between bands $l$, $n$, and $\omega_{nl} = \omega_{n\mathbf{k}} - \omega_{l\mathbf{k}}$ is the resonant frequency between bands. We let $V_\mathrm{u.c.}$ be the unit cell volume.

We briefly comment on the significance of the the second term in Eq.~\eqref{eq:free-energy}, that is, the charging of the dielectric, insulating medium,
\begin{align}
 -&\mathbf{E} \cdot \mathbf{p}(\mathbf{u},\mathbf{E}) = -\frac{V_\mathrm{u.c.}\epsilon_0}{2} \sum_{ab} \Re\left(\chi^{ab}(\omega) E_a E_b^*\right).
    \label{eq:field-energy}
\end{align}

This term however is subleading in determining the energy change of the free energy. 
Assuming a characteristic density of states in the valence band $D$, velocity $v$ and relaxation time $\gamma_{ln}^{-1}\sim\tau$, the occupation energy change expressed through Eqs.~\eqref{eq:band}-\eqref{eq:density_matrix} takes the order of magnitude given by $\omega\tau\frac{e^2 v^2 V_\mathrm{u.c.} D E^2}{\hbar\omega^2}$, while the charging energy Eq.~\eqref{eq:field-energy} if of the order $\frac{e^2 v^2 V_\mathrm{u.c.} D E^2}{\hbar\omega^2}$. As the ratio between them is $\sim\omega\tau \gg 1$, we safely neglect corrections from dielectric charging (Eq.~\eqref{eq:field-energy}). Finally, combining Eqs.~\eqref{eq:band}-\eqref{eq:density_matrix}, the total change in the free energy (per unit cell) can be expressed as,
\begin{align*}
    &G(\mathbf{u},\mathbf{E}) - G_0(\mathbf{u})\\
    =&  \frac{i e^2V_\mathrm{u.c.}}{\hbar \omega^2} \sum_{nlab} 
    \int_\mathbf{k}    \frac{f_{ln} \omega_{nl} v_{nl}^a v_{ln}^b }{\gamma_{nl}(\omega_{ln} - \omega + i\gamma_{ln})} E_a(\omega) E_b(-\omega) 
    \numberthis
    \label{eq:final-expr}
\end{align*}
Eq.~\eqref{eq:final-expr} is the main result of our work, and enables the calculation of the correction to the free energy for a given ionic configuration $\mathbf{u}$ under an applied electric field $\mathbf{E}$ using the static, equilibrium Bloch states $n,l$. 

We note several features of Eq.~\eqref{eq:final-expr}. First, this response function defined by a rank-2 symmetric tensor $\tilde{\chi}^{ab}_\mathrm{ex}=\frac{\partial^2 G(\mathbf{u},\mathbf{E})}{\partial E_a \partial E_b^*}$ is symmetric (even) under time-reversal, spatial inversion, and complex conjugation, consistent with physical constraints (see SI) and Kramers-Kronig relations. It is generically non-vanishing for any point group symmetry. Second, the numerator in Eq.~\eqref{eq:final-expr} is proportional to matrix elements of Bloch velocity operators, which encode the quantum geometric properties of the solid \cite{Ahn_2021,Kaplan2023}. Third, adopting the simple relaxation time approximation $\gamma_{ln} = \tau^{-1}$ independent of band energy or momentum, Eq.~\eqref{eq:final-expr} is related to a fluctuation-dissipation relation, connecting this nonlinear response with the linear optical conductivity (see SI). 

\paragraph{Multiferroic transition in bilayer CrI\textsubscript{3} and T\textsubscript{d}-MoTe\textsubscript{2}. ---}
Eq.~\eqref{eq:final-expr} allows for the calculation of the correction to the free-energy in the presence of arbitrary symmetry. We now show that in the presence of local minima in the free energy, -- such as those present in ferroelectric or multiferroics (with concomitant magnetic order) -- Eq.~\eqref{eq:final-expr} enables switching of order. 

To construct the bilayer, we stack monolayers such that inversion symmetry is not restored when both are within the same unit cell. This can be done in the vdW/TMD context by applying the $M_z$ operation on one layer, consistent with the prescription in Ref.~\cite{ji_general_2023} that enables ferroelectricity.
A stacking configuration $\mathbf{u}$ is then formed by shifting one layer with respect to another by $(x,y)\to(x+a u_x,y+au_y)$ where $a$ is the lattice constant.
Such structures, as in the case of hBN and CrI\textsubscript{3} \cite{Lei_2021} have been realized experimentally.

We now show that Eq.~\eqref{eq:final-expr} allows for switching between different ferroic orders. We used the PBE-GGA functional \cite{perdew_gga_1996} within DFT as implemented in the Vienna ab-initio simulation package (VASP) \cite{kresse1996efficiency,Kresse1996Efficient} to compute the free energy $G(\mathbf{u},0)$ for every stacking $\mathbf{u}$. The free energy landscape for CrI\textsubscript{3} is plotted in Fig.~\ref{fig:2}(a). Consistent with the conjectured picture of a large ground state degeneracy, we find minima separated by energy barriers. Of these, we extract an MEP, which we highlight by an orange line on Fig.~\ref{fig:2}(a) (see also Fig.~\ref{fig:4}(c)). 

We then calculate Eq.~\eqref{eq:final-expr} by means of a Wannier interpolation of electronic states for every stacking $\mathbf{u}$~\cite{Marzari2012,pizzi2020wannier90}. We scan the applied frequency $\omega$ for the frequency dependence of the free energy $G(\mathbf{u},\mathbf{E})$ at $A$, $S$, and $B$ stackings as shown in Fig.~\ref{fig:2}(b). For two frequencies which bias $A \to B$ or $B \to A$ we plot the free energy $G(\mathbf{u}, \mathbf{E})$ as a function of light intensity, $\sim \mathbf{E}^2$. 

In Fig.~\ref{fig:2}(c) we consider the transition from $A \to B$ ($\hbar\omega=3.45\mathrm{eV}$) while in Fig.~\ref{fig:2}(d) we have $B \to A$ ($\hbar\omega=2.55\mathrm{eV}$). In all cases, we find a critical electric field strength ($I_c=6\sim10\text{GW/cm}^2$) at which switching occurs which is evident by the softening of the minimum at a given stacking $A,B$. This is concretely plotted in Fig.~\ref{fig:2}(e,f), where we show the disappearance of the minima near $A,B$ (depending on the light frequency) along the MEP, as a function of intensity. A qualitatively similar picture is found for MoTe\textsubscript{2}, whose ground state stacking energy is plotted in Fig.~\ref{fig:4}(c). We use MoTe\textsubscript{2} to showcase the utility of our method for switching stackings $A,B$, which are not necessarily global minima of the bilayer structure. In MoTe\textsubscript{2}, $A, B$ are significantly inequivalent, compared with \cri. Our results clearly show the potential of our method to switch as shown in Fig.~\ref{fig:4}(e,f). We also identified a much richer polarization landscape in bilayer MoTe\textsubscript{2} as shown in  Fig.~\ref{fig:4}(d), owing to its lower point group symmetry. For \mztmote the MEP is spanned by the stackings $A(\frac{1}{2}, 0)$ and $B(\frac{1}{4}, \frac{1}{2})$ (Fig.~\ref{fig:4}(c)), accompanied with out-of-plane polarization $P_z=0$ and $P_z=10.5\times10^{-3} \textrm{eÅ}/\textrm{u.c.}$ respectively (Fig.~\ref{fig:4}(d)). 
{\allowdisplaybreaks
\begin{figure}[th]
\includegraphics[width=\linewidth, page=2]{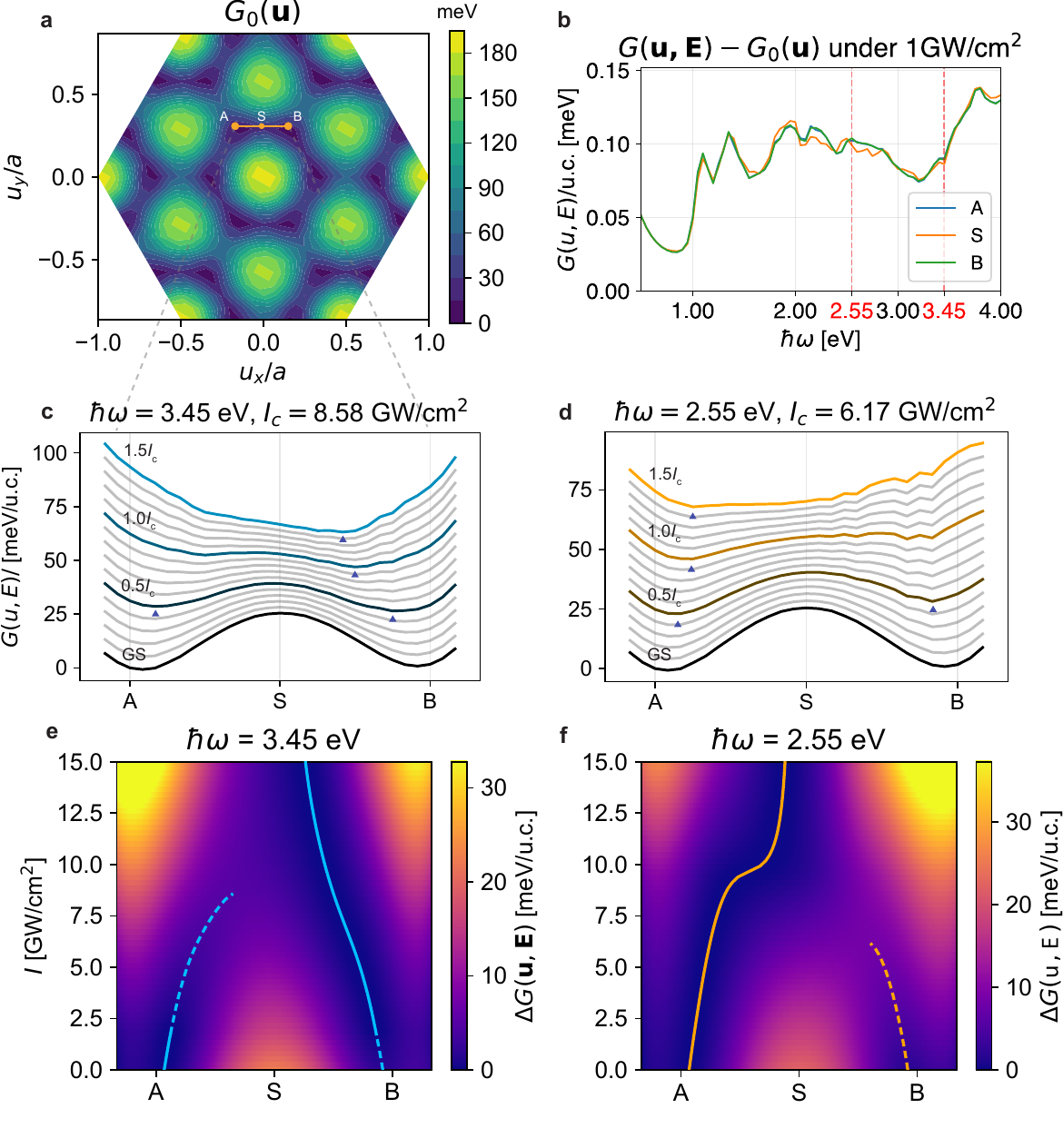}  
\caption{\label{fig:2} Light-induced sliding in bilayer \mzcri. (a) Equilibrium energy of bilayer \mzcri of different stackings parameterized by $\mathbf{u}=(u_x, u_y)$, the in-plane relative translation vector of one layer. Indicated as a yellow line is a minimum energy path (MEP) between two locally stable stackings $A$ and $B$, through a saddle point $S$. (b) the nonlinear excitation free energy under 1GW/cm\textsuperscript{2} circularly polarized light in the three key stackings $A$, $S$, and $B$. Small variations in the excitation energies are utilized to induce sliding. (c)(d) As pumping light intensity increases from 0 (ground state, GS) to 0.5, 1, 1.5 times the respective critical value, a structural transition in both directions can be induced at 3.45 eV (c, from $A$ to $B$, $\lambda=$359nm) and 2.55 eV (d, from $B$ to $A$ $\lambda=$486nm). (e)(f) Under the two illumination conditions in (c) and (d), color map shows the excitation free energy relative to the minimal energy at the same intensity among stackings, $G(\mathbf{u}, \mathbf{E}) - \min_\mathbf{u} G(\mathbf{u}, \mathbf{E})$. Curves show the locally stable stackings, with globally stable points in solid lines and metastable points in dashed lines.}
\end{figure}}

{\allowdisplaybreaks
The controllable light-induced sliding, or ferroelastic switching, is also accompanied by a ferroelectric switching, because the bilayer vdWs exhibit different strengths and directions of polarization, as shown in Fig.~\ref{fig:3}(a) and \ref{fig:4}(d). More broadly, several symmetries control the polarization between stackings \cite{ji_general_2023}: $P(C_{2z}\mathbf{u})=-P(\mathbf{u})$, $P(m_{110}\mathbf{u}) = P(\mathbf{u})$, and $P(m_{1\bar{1}0}\mathbf{u}) = -P(\mathbf{u})$, thus forming an alternating pattern in the sliding position space. Importantly, the two locally stable positions $A$ and $B$ have opposite polarizations, so sliding is necessarily accompanied by an out-of-plane ferroelectric dipole flip.

Independent from the two ferroic orders discussed above, there is a concomitant nonlinear Edelstein effect (NLEE) \cite{xu_light-induced_2021,zhang_all-optical_2022} at all stackings, which will be responsible for the magnetization reversal. We computed the NLEE effect in \mzcri at stacking $A$ using the same method as in Ref.~\cite{xu_light-induced_2021}, shown in Fig.~\ref{fig:3}(c). At frequencies between 1.0 and 2.7 eV, light intensity above $\sim$ 33GW/cm\textsuperscript{2} can induce magnetization comparable to the original magnetization $\sim \mu_B$, and thereby induce magnetic switching. The direction of induced magnetization will depend on the chirality of light. This establishes our effect as genuinely multiferroic.

\begin{figure}[th]
\includegraphics[width=\linewidth, page=3]{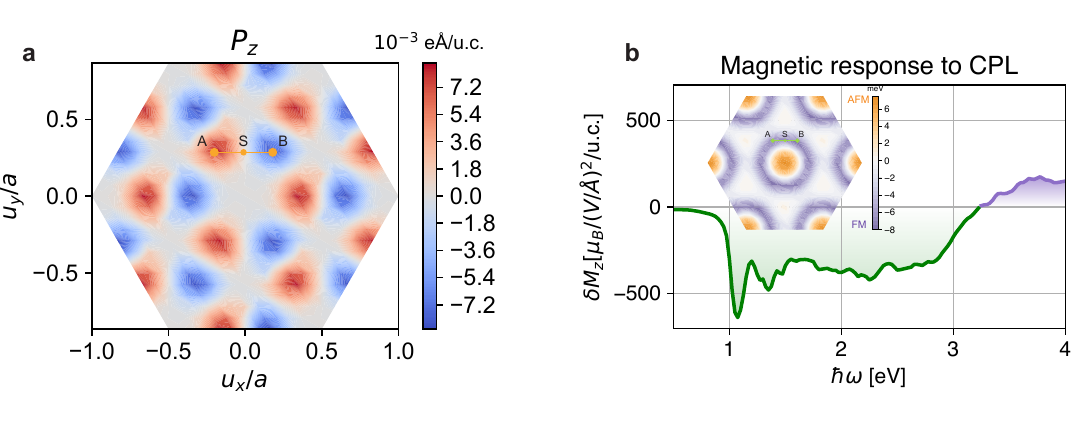}     
\caption{\label{fig:3} Multiferroic transition induced by nonlinear interaction. (a) Out-of-plane stacking-dependent ferroelectric polarization dipole in the ground state. Also showing the same transition path $A-S-B$ marked in Fig.~\ref{fig:2}(a). The alternating polarization pattern allows ferroelectric order flipping along with the structural transition. (b) At stacking $A$, the magnetization generated by Nonlinear Edelstein Effect (NLEE) as a function of incident circularly-polarized photon energy. Inset shows the DFT energy difference $E_{FM}-E_{AFM}$, indicating that ferromagnetic order is preferable throughout the transition path $A-S-B$. }
\end{figure}
}
\paragraph{Quantum geometry of light-induced sliding.---}
\allowdisplaybreaks
As shown, Eq.~\eqref{eq:final-expr} is intimately connected to the matrix elements of Bloch states, defining quantum geometry \cite{torma2023essay,liu2024quantum,yu2024quantum,ma2021topology}. We use the property that $v^a_{nl} = i\omega_{nl}r_{nl}^a$, where $r_{nl}^a = \langle u_{n\mathbf{k}}|\partial_{k_a}|u_{l\mathbf{k}}\rangle$ is the relation between the velocity operator and the generally (non-Abelian) position operator.  Working in the relaxation time approximation and assuming $\omega\tau\gg 1$, $\omega_{nl} = \omega$. We recast Eq.~\eqref{eq:final-expr} as:
\begin{align}
    \frac{\partial^2 G(\mathbf{u},\mathbf{E})}{\partial E_a(\omega) \partial E_b(-\omega)} = \frac{V_\mathrm{u.c.}\pi e^2 \omega\tau}{\hbar} \sum_{nl}\int_\mathbf{k} f_{nl}r^a_{nl}r^b_{ln}\delta(\omega-\omega_{ln})
    \label{eq:simplified}
\end{align}
We recognize the quantity $r^a_{nl}r^b_{ln}$ as the quantum geometric tensor $Q^{ab}_{nl} = g^{ab}_{nl} + i\Omega^{ab}_{nl}/2$ \cite{cheng2013quantumgeometrictensorfubinistudy}. Here $g^{ab}_{nl}$  is the (non-Abelian) quantum metric for bands $n,l$ while $\Omega^{ab}$ is their Berry curvature. Unsurprisingly, this quantity governs the entirety of the coupling to the electric field, as argued in the past \cite{holder2020consequences, Kaplan2023}, since $g^{ab}$ is  symmetric in $a,b$ it is primarily responsible for linear-polarized light coupling while $\Omega^{ab}$ being antisymmetric couples only to circularly polarized light. However, it is $\Omega^{ab}$ that controls the induction of a magnetic moment in the NLEE \cite{xu_light-induced_2021}, thus unifying the ionic transition with the electronic one. We note an important consequence of this observation: Ref.~\cite{verma2025instantaneousresponsequantumgeometry} introduced the time-dependent quantum geometric tensor (tQGT) as,
\begin{align}
    Q^{ab}(t-t')= \sum_{nl} f_n(1-f_l)e^{i\omega_{nl}(t-t')} r^a_{nl}r^b_{ln}.
    \label{eq:QGT}
\end{align}
Following Ref.~\cite{verma2025instantaneousresponsequantumgeometry}, the physically relevant quantity is the antisymmetric part of this tensor, i.e., $Q_{\textrm{asym.}}^{ab}(t) = (Q^{ab}(t)-Q^{ba}(-t))/2$. 

Computing the $\omega$ harmonic of $Q_{\textrm{asym.}}^{ab}(t)$ we find, for insulators, (setting $t-t'\to t$),

\begin{align}
    \notag Q^{ab}_{\textrm{asym}.}(\omega)&= \sum_{nl} f_n(1-f_l)\biggl[\delta(\omega-\omega_{ln})r^a_{nl}r^b_{ln}- \\ & \notag \delta(\omega-\omega_{nl})r^b_{nl}r^a_{ln}\biggr] = \\ & \sum_{nl} f_{nl} \delta(\omega-\omega_{ln})r^a_{nl}r^b_{ln},
\end{align}
identical with the matrix elements appearing in Eq.~\eqref{eq:simplified}.
Likewise, we note that for NLEE \cite{xu_light-induced_2021}, the response tensor is equal to $\xi^{z}_{a,b} = \pi \tau \frac{e^2}{2\hbar^2}\int_{\mathbf{k}}f_{nl}(r^a_{nl}r^b_{lm}-r^b_{nl}r^a_{ln})(L_n-L_l)\delta(\omega-\omega_{nl})$. For bands of well-defined angular momentum (as in \cri), We replace $L_n-L_l$ with a constant $\Delta L$, which leads to $\xi^{z}_{a,b} = \pi \tau \frac{e^2 \Delta L}{2\hbar^2}\int_{\mathbf{k}}\textrm{Im}(Q^{ab}_{\textrm{asym.}}(\omega))$.
This establishes light-induced Fermi excitations, Eq.~\eqref{eq:density_matrix}, as the frequency harmonics of the tQGT, with the light-induced structural and magnetic transitions forming a sensitive, resonant probe of quantum geometry. As shown in Fig.\ref{fig:excited_bandstructure}, the transitions between states which enable the sliding in \cri occur between Cr orbitals of varying angular quantum number $m$. We note that similar polarization generation was recently proposed for superconductors with broken inversion symmetry \cite{Kaplan2025quantum}. The change in electronic occupation is itself related to gravitational anomalies in solids, closely connected with quantum geometry~\cite{mitscherling2020longitudinal,holder2021mixed}.
\begin{figure}[th]
\includegraphics[width=\linewidth, page=4]{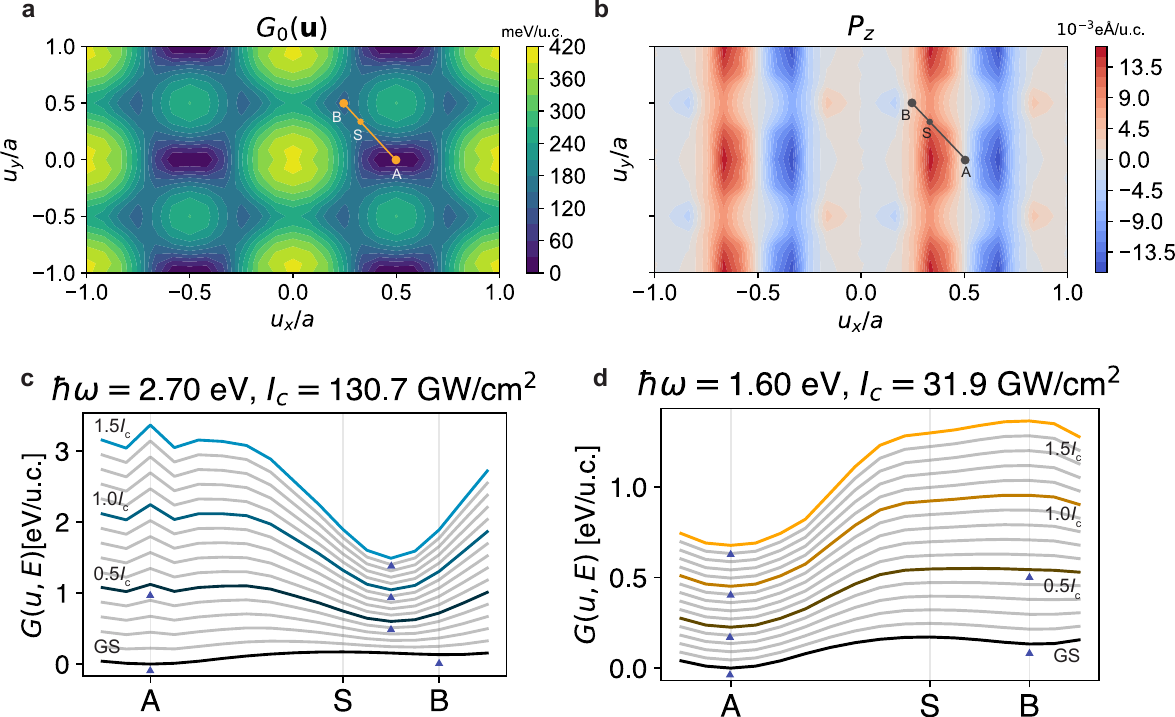}     
\caption{\label{fig:4} Light-induced sliding in bilayer \mztmote. (a) Ground state energy and (b) out-of-plane polarization of \mztmote, with transition path $A-S-B$ indicated. (c)(d) Excited state energy under two different frequencies of pumping light, used to induce (c) $A-B$ transition with 2.7eV photon energy ($\lambda=459$nm) and (d) $B-A$ transition with 1.6eV photon energy ($\lambda=774$nm). Curves show free energy as light intensities are increased from 0 (ground state, GS) to 0.5, 1.0, 1.5 times the respective critical intensity for transition. }
\end{figure}

\paragraph{Discussion.---}

In this work, we demonstrated how nonlinear light-matter interactions leads to structural and electronic transitions. By employing first-principle calculations we mapped out the possible stacking configurations of bilayer \cri and \mote, showing a rich landscape of polarization and magnetization that can be tuned. We then computed our reported effect, the NELIC, which causes layer sliding in \cri and \mote, as well as inducing an AFM to FM transition (in the case of \cri). This establishes the NELIC as a versatile, adjustable and readily generalizable tool for the in-situ manipulation of order in quantum materials.

Several works \cite{zhang_light-induced_2020,yang2024_lightinduced,Mankowsky_2017,nova2017effective,Gao2024}  have already pointed out the possibility of manipulating electronic phases (whether polar or magnetic) with light, by coupling to phonons. Our approach permits extending this paradigm to optical pulses which can be localized to parts of the sample, and engineering domains \cite{rubio2015ferroelectric} can then be realized on the sub-micron scale. New promising directions include the extension of our results to the recently discovered spatiotemporal order in solids \cite{Kaplan_2025_spatiotemporal}, and subgap phenomena involving multiferroicity \cite{shi2024nonresonantramancontrolmaterial,kaplan2020nonvanishing,kaplan2023general}. Recently, a free-energy mechanism has been shown to produce N\'eel-vector switching in $\mathcal{P}\mathcal{T}$ symmetric AFMs~\cite{xue2025reversingneelvectorptantiferromagnets}.

The quantum geometry of bands will serve as an important design principle for light-induced structural transitions; as previously argued \cite{komissarov2024quantum}, covalent versus ionic bonding significantly impacts the quantum metric contribution to response functions. We anticipate that bond engineering \cite{huang2025understanding}, or symmetry breaking \cite{mei2024electrical} in multilayered vdW structures will significantly enhance our proposed mechanism.

Overall, our framework serves as groundwork for utilizing nonlinear optical control of electronic excited states to achieve fast, efficient, and reversible structural and multiferroic order switching, with broad implications for future optoelectronic and nanophotonic technologies.

\paragraph{Acknowledgments.---}

    We are grateful to Premala Chandra, Cristian D Batista, Riccardo Comin, Gaurav Chaudhary, Enrico Rossi, Nish Verma, Raquel Queiroz, Yafis Barlas, Benjamin Remez and Daniel Bennett for stimulating discussions. 
    H.W. is supported by Senior Undergraduate Research Fellowship, TEEP, Tsinghua University. 
    D.K. is supported by the Abrahams Postdoctoral Fellowship of the Center for Materials Theory, Rutgers University, and the Zuckerman STEM Fellowship.
    D.K. acknowledges the hospitality of the Aspen Center for Physics, supported by National Science Foundation grant PHY-2210452 as well as partial support by grant NSF PHY-2309135 to the Kavli Institute for Theoretical Physics (KITP), where part of this work was carried out. J.L. acknowledges support by DTRA (Award No. HDTRA1-20-2-0002) Interaction of Ionizing Radiation with Matter (IIRM) University Research Alliance (URA).

\bibliography{main.bbl}  
\clearpage
\newpage
\onecolumngrid
\begin{center}
{\bf{Supplementary Information: Ultrafast switchable polar and magnetic orders by nonlinear light-matter interaction}}\\
{{ Haoyu Wei, Daniel Kaplan, Haowei Xu, and Ju Li}}
\end{center}

\renewcommand\thesection{\arabic{section}}
\renewcommand\theequation{S\arabic{equation}}
\renewcommand\thefigure{S\arabic{figure}}

\setcounter{equation}{0}
\section{Excitations from time-dependent Density Functional Theory}

In time-dependent density functional theory (TDDFT) \cite{marques_fundamentals_2012}, the time-dependent Kohn-Sham (KS) equations can be formulated as,
\begin{equation}
    i \frac{\partial \phi_{j}(\mathbf{r}, t)}{\partial t} = \left[-\frac{1}{2} \nabla^2 + {v}_\mathrm{KS}[n; \Phi_0](\mathbf{r}, t)\right] \phi_j(\mathbf{r}, t)
    \label{eq:TDKS}
\end{equation}
where $\phi_j$ are the time-dependent KS orbitals, and $\Phi_0$ is the initial many-body state e.g. a Slater determinant of single-particle orbitals $\phi_j(r)$. ${v}_\mathrm{KS}$ is the KS single-particle effective potential composed of three terms
\begin{equation}
   {v}_\mathrm{KS}[n; \Phi_0](\mathbf{r}, t) = {v}_\mathrm{ext}[n; \Psi_0](\mathbf{r}, t) + e^2\int \dd^3 \mathbf{r}' \frac{n(\mathbf{r}', t)}{|\mathbf{r} - \mathbf{r}'|} + v_\mathrm{xc}[n; \Psi_0, \Phi_0](\mathbf{r}, t)
\end{equation}
where $v_\mathrm{ext}[n;\Psi_0](\mathbf{r},t)$ is the external time-dependent field, and the third term $v_\mathrm{xc}$ is the exchange-correlation (xc) potential, defined to be the difference between the potential that generates density $n(\mathbf{r},t)$ in an interacting system starting in initial state $\Psi_0$ and the single-particle potential that generates
this same density in a non-interacting system starting in initial state $\Phi_0$. In the small excitation regime, the exchange-correlation functional $E_\mathrm{xc}[n(\mathbf{r})]$ and thus $v_\mathrm{xc}[n(\mathbf{r})]$ are empirically taken to be the same form as the ground state \cite{marques_fundamentals_2012}. Similar to static DFT, the energy of the excited state can be written as a function of the KS orbitals
\begin{equation}
    \expval{H_\mathrm{ex}} = \sum_j \rho_j \mathcal{E}_j + E_\mathrm{xc} [n(\mathbf{r})] 
    - \int {v}_\mathrm{xc}(\mathbf{r}) n(\mathbf{r}) \dd \mathbf{r}
    - \frac{e^2}{2} \iint \frac{n(\mathbf{r})n(\mathbf{r}')}{|\mathbf{r} - \mathbf{r}'|} \dd^3\mathbf{r} \dd^3\mathbf{r}'
    \label{eq:TD-energy}
\end{equation}
where $\mathcal{E}_j$ and $\rho_j$ are the energy and occupation number of the $j$-th orbital. Since the density perturbation $\Delta n(\mathbf{r})$ is a small quantity in the perturbative regime, the increase in the many-body total energy Eq.~\eqref{eq:TD-energy} can be expanded as
\begin{align}
   \expval{H_\mathrm{ex}} - \expval{H_\mathrm{gr}} = & \sum_j \mathcal{E}_j \Delta \rho_j + \sum_j \Delta\mathcal{E}_j  \rho_j + \int\left(\frac{\delta E_\mathrm{xc}[n(\mathbf{r})]}{\delta n(\mathbf{r})}\right) \Delta n(\mathbf{r}) \dd^3\mathbf{r} \nonumber \\
   & - \int v_\mathrm{xc}(\mathbf{r}) \Delta n(\mathbf{r}) \dd^3 \mathbf{r} - \iint \frac{n(\mathbf{r})\Delta n(\mathbf{r})}{|\mathbf{r} - \mathbf{r}'|} \dd^3 \mathbf{r} \dd^3 \mathbf{r}' 
   \label{eq:ex-energy}
\end{align}
The third and fourth term on the right hand side of Eq.~\eqref{eq:ex-energy} will cancel. Also, in our case, the second term cancels out with the last term. 
To see the latter fact, we need to evaluate the shift of eigenenergies $\Delta \mathcal{E}_j$ due to light excitation. It can be obtained by applying a first-order perturbation on the TDKS eigenequation Eq.~\eqref{eq:TDKS}. The result is
\begin{equation}
    \Delta \mathcal{E}_j = \langle \phi_j | \Delta H_\mathrm{KS} | \phi_j \rangle = \int \dd^3 \mathbf{r} \phi_j^{*} v_\mathrm{ext} \phi_j + \iint \frac{n(\mathbf{r}')\Delta n_j(\mathbf{r})}{|\mathbf{r} - \mathbf{r}'|}  \dd^3 \mathbf{r} \dd^3 \mathbf{r}'
    \label{eq:3.16}
\end{equation}
where the first term on the right hand side of Eq.~\eqref{eq:3.16} is the energy induced by the external field on the $j$-th KS orbit. In the present scenario where $v_\mathrm{ext}$ is the oscillating optical electric field potential, its time average will be zero on a timescale relevant for ionic transitions. Ignoring this oscillating term, the second term in Eq.~\eqref{eq:3.16} cancels the last term in Eq.~\eqref{eq:ex-energy}. Summing up the rest, the final form of Eq.~\eqref{eq:ex-energy} becomes
\begin{equation}
    \expval{H_\mathrm{ex}} - \expval{H_\mathrm{gr}} = \sum_j \mathcal{E}_j \Delta \rho_j
\end{equation}
i.e., the excitation energy of the many-body system is equal to the excitation energy of the KS system. This is Eq.~\eqref{eq:band-theory} in the main text. 

\section{Derivation of the excitation energy Eq. (7)}
In a quantum system with known energy levels $\mathcal{E}_m$, we use density matrix perturbation theory Ref.~\cite{boyd_nonlinear_2019}, with a perturbative potential $\hat{V}$ of the form
\begin{equation}
    \hat{V} = \sum_p \hat{V}(\omega_p) e^{-i\omega_p t},
\end{equation}
the second-order density matrix response is
\begin{align*}
    \rho_{nm}^{(2)} = \sum_{\nu} \sum_{pq} & e^{-i(\omega_p + \omega_q)t}\\
    \times &\left\{
    \frac{\rho_{mm}^{(0)} - \rho_{\nu\nu}^{(0)}}{\hbar^2}
    \frac{V_{n\nu}(\omega_q)V_{\nu m} (\omega_p)}
    {[(\omega_{nm} - \omega_p - \omega_q) - i\gamma_{nm}][(\omega_{\nu m} - \omega_p) - i\gamma_{\nu m}]}\right.\\
    -&
    \left.
    \frac{\rho_{\nu\nu}^{(0)} - \rho_{nn}^{(0)}}{\hbar^2}
    \frac{V_{n\nu}(\omega_p)V_{\nu m} (\omega_q)}
    {[(\omega_{nm} - \omega_p - \omega_q) - i\gamma_{nm}][(\omega_{n\nu} - \omega_p) - i\gamma_{n\nu}]}
    \right\}
    \numberthis
\end{align*}
where $\rho^{(r)}_{nm}$ is the $r$-th order perturbation in the density matrix evaluated on its $n,m$-th element. The $\omega$ with double indices e.g., $\omega_{mn}=\frac{\mathcal{E}_m-\mathcal{E}_n}{\hbar}$ is the frequency gap between $m$ and $n$ levels, same as in the main text; while those with single indices e.g., $\omega_p$ are frequencies of the perturbative potential $\hat{V}$. 

In our case, only the DC component ($\omega_p + \omega_q = 0$) of the diagonal ($n=m$) elements are relevant. Using the ground initial state condition $f_n = \rho_{nn}^{(0)}$, and the light-matter coupling Hamiltonian $V_{mn}(\omega) = -e{\mathbf{v}}_{mn}\cdot \mathbf{A} = \frac{ie}{\omega} \mathbf{v}_{mn}\cdot \mathbf{E}$, the diagonal elements of the DC density matrix, to the first non-vanishing order, is
\begin{align}
    \rho_{nn}^{(2)}
= -\frac{i \tau e^2}{\hbar^2\omega^2} \sum_{nlab\omega} \left(  \frac{f_{nl} v_{nl}^a v_{l n}^b }{\omega_{l n} - \omega + i\gamma_{ln}} 
+ \frac{f_{nl} v_{l n}^a v_{nl }^b }{\omega_{nl } - \omega + i\gamma_{nl}} \right) 
E_a(\omega) E_b(-\omega)
\end{align}

Therefore, the excitation energy 
\begin{align}
\expval{H_\mathrm{ex}} - \expval{H_\mathrm{gr}}
&= \sum_n \rho_{nn}^{(2)} \hbar \omega_n \\
&= -\frac{i \tau e^2}{\hbar\omega^2} \sum_{nlab\omega} \left(  \frac{f_{nl} v_{nl}^a v_{l n}^b \omega_n}{\omega_{l n} - \omega + i\gamma_{ln}} 
+ \frac{f_{nl} v_{l n}^a v_{nl }^b \omega_n }{\omega_{nl } - \omega + i\gamma_{nl}} \right) 
E_a(\omega) E_b(-\omega) \\
&= -\frac{i \tau e^2}{\hbar\omega^2} \sum_{nlab\omega} \left(  \frac{f_{nl } v_{nl }^a v_{l n}^b \omega_n}{\omega_{l n} - \omega + i\gamma_{ln}} 
- \frac{f_{nl} v_{n l}^a v_{l n}^b \omega_l  }{\omega_{l n} - \omega + i\gamma_{ln}} \right) 
E_a(\omega) E_b(-\omega) \\
&= -\frac{i \tau e^2}{\hbar\omega^2} \sum_{nlab\omega} \frac{f_{nl } v_{nl }^a v_{l n}^b \omega_{nl}}{\omega_{l n} - \omega + i\gamma_{ln}} 
E_a(\omega) E_b(-\omega) 
\end{align}

In our case of Bloch states, the states are also labeled by their $\mathbf{k}$ vectors. The NELIC excitation energy, after ignoring the $\mathbf{E}\cdot \mathbf{p}$ term, is given by
\begin{align}
    {G} - G_0 = \frac{i \tau e^2 V_\mathrm{u.c.}}{\hbar\omega^2} \sum_{nl\omega} \int_{\mathbf{k}} \frac{f_{nl }\omega_{ln} v_{nl }^a v_{l n}^b }{\omega_{l n} - \omega + i\gamma_{ln}} 
E_a(\omega) E_b(-\omega),
\end{align}
where as usual, $\int_{\mathbf{k}} (\cdot) = \int \frac{\dd^d \mathbf{k}}{(2\pi)^d} (\cdot)$, and all electronic variables are implicitly labeled by $\mathbf{k}$ in addition to the explicit band indices. 

And the corresponding response function $\tilde{\chi}_{\textrm{ex}}^{ab}(\omega)$, defined by $G-G_0 = \sum_{ab,\omega} \tilde{\chi}_{\textrm{ex}}^{ab}(\omega) E_a(\omega) E_b(-\omega)$, is given by 
\begin{align}
    \tilde{\chi}_{\textrm{ex}}^{ab}(\omega) = \frac{i \tau e^2 V_\mathrm{u.c.}}{\hbar\omega^2} \sum_{nl\omega} \int_{\mathbf{k}} \frac{f_{nl }\omega_{ln} v_{nl }^a v_{l n}^b }{\omega_{l n} - \omega + i\gamma_{ln}} 
    \label{eq:chi,8}
\end{align}

\section{Fluctuation-dissipation relation in the uniform decay rate $\gamma$ limit} 
In the constant decay rate limit, the steady state of the excited electron state is one where the rate of excitation equals the rate of decay, so are the rate of energy excitation and dissipation. Therefore, the steady-state excitation energy $\Delta G$ equals the energy dissipation rate times the excitation lifetime $\tau = {\gamma}^{-1}$ (note that we can only do so in the case where decay rates are equal for all energy levels, i.e. uniform $\gamma$). On the other hand, the energy dissipation comes from Joule heating, and the dissipation rate would be $\mathbf{J}(\omega)\cdot \mathbf{E}^*(\omega)=\sum_{ab}\sigma^{ab}(\omega)E_a(\omega) E_b(-\omega)$, where $\sigma^{ab}(\omega)$ is the optical conductivity at frequency $\omega$. Therefore we find $\Delta G = \tau \times V_\mathrm{u.c.}\sum_{ab\omega} \sigma^{ab}(\omega)E_a(\omega) E_b(-\omega)$. Plugging in the Kubo formula for optical conductivity\cite{ashcroft_solid_1976,komissarov2024quantum}
\begin{equation}
    {\sigma }^{ab }(\omega )=-\frac{i{e}^{2}}{\hslash }\mathop{\sum}_{n\ne m} \int_\mathbf{k}
    \frac{{f}_{nm}{\omega }_{nm}{r}_{nm}^{a}{r}_{mn}^{b}}{{\omega }_{nm}+\omega +i\tau^{-1}},
\end{equation}
we get an expression for $\Delta G$ that agrees with Eq.~\eqref{eq:final-expr} in Main Text assuming uniform $\gamma$.

\section{Existence of switching behavior from a symmetry perspective}
For switching behavior to be present in a 2D material under this mechanism, the material must have two stable states that have nonequivalent free-energy response under light. Crucially, we want to rule out identical response in these stable stackings constrained by intrinsic crystal symmetries. 

If a bilayer is composed of identical single-layers with in-plane mirror symmetry $M_z$, its two opposite stackings $\mathbf{u}$ and $-\mathbf{u}$  behave identically to NELIC. Because in this case the stackings $\pm\mathbf{u}$ are related by a global $M_z$, both ground state energy and excited state energy Eq.~\eqref{eq:final-expr} are invariant under $M_z$, making them respond degenerately under NELIC. 

\begin{figure}[ht]
    \centering
    \includegraphics[width=0.8\linewidth, page=6]{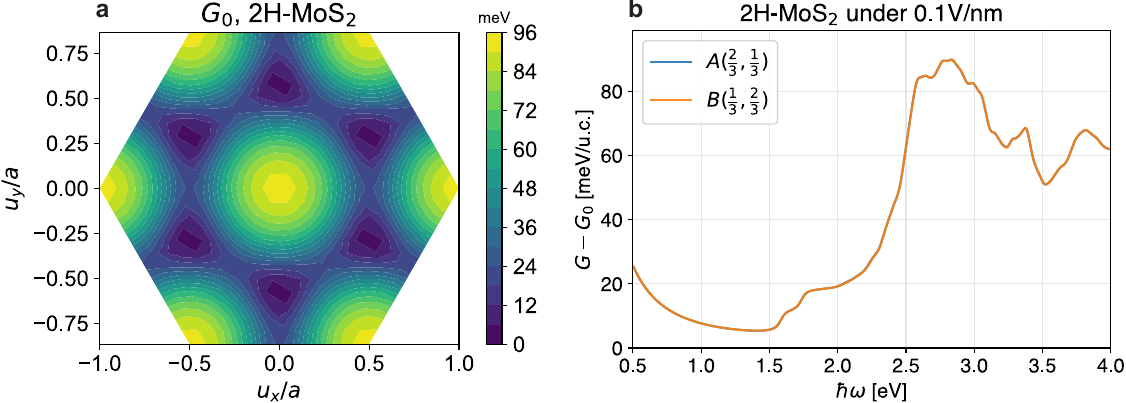}
    \caption{(a) Ground state energy landscape of 2H-MoS\textsubscript{2} at different stackings. (b) Excitation energy response to 2H-MoS\textsubscript{2} at the two stable stackings $A$ and $B$. Linearly polarized light along y direction is used for this plot, but we see similar bahavior in other polarizations as well. }
    \label{fig:2h-mos2}
\end{figure}

Common choices such as 2H phases of TMD, and planar layers such as graphene and BN fail to satisfy the criterion of an optically controlled sliding switch because of this symmetry rule. In Fig.~\ref{fig:2h-mos2}, we show bilayer 2H-MoS\textsubscript{2} as an example. The two translationally nonequivalent stackings $A(\frac{2}{3}, \frac13)$ and $B(\frac13,\frac23)$ (a) are related by a global $M_z$ symmetry, therefore the response are identical and unsuitable for light-induced switching. 

\section{Stacking symmetries in \cri}
As the symmetry in \mzcri can explain our choice of the structural switching path and the presence of alternating that enables ferroelectric switching, we provide a detailed analysis of the symmetry in ground state \mzcri here.

The ground state energy for stackings marked by relative interlayer sliding $G_0(\mathbf{u})$, shown in Fig 2 (a), apparently has inversion symmetry $C_{2z}$ and mirror symmetries $m_{110}$ and $m_{1\bar{1}0}$
To understand this fact, here we introduce an analytical approach to deriving the symmetries. We denote the set of all atomic positions in a single \cri layer by $L_1 = \{(x_i, y_i, z_i; n_i)\}$, the collection of atomic positions ($x_i,y_i,z_i$) and type (marked by $n_i$, which can be its atomic number). 
Then, the set of atoms in a \textit{bilayer} with sliding vector $\mathbf{u}$ would be 
\begin{equation}
    L_2(\mathbf{u}) = T(\mathbf{u}) M_z L_1 \bigcup L_1
    \label{eq:mzcri-bilayer-set}
\end{equation}
where $T(\mathbf{u})$ is a translation operator that displaces the atomic set $L_1$ by $\mathbf{u}$, and $M_z$ is a mirror symmetry operator with respect to a mirror plane perpendicular to $\hat{z}$) halfway between the two layers. Therefore, $T(\mathbf{u}) m_z L_1$ would mean that $L_1$ is mirrored first (positioned at the correct $z$ coordinate at the same time), and then translated in the $x-y$ plane by $\mathbf{u}$, to make the second layer. This second layer is then combined ("$\cup$") with the first layer $L_1$ to make the bilayer system, denoted as $L_2(\mathbf{u})$.

With this tool, we can analyze a few symmetries the ground state energy $G_0 (\mathbf{u})$, and the excited state response function $\tilde{\chi}_\mathrm{ex}^{ab}$ also has properties following from these stacking symmetries. 

\paragraph{$C_{2z}$ symmetry of the bilayer.} 
If we apply a global $T(-\mathbf{u}) m_z$ operation to $L_2(\mathbf{u})$, then the resulting structure should have the same ground state energy with the initial $L_2(\mathbf{u})$. On the other hand, the new system under operation is
\begin{equation}
    L_2' = T(-\mathbf{u}) m_z L_2(\mathbf{u}) = L_1 \bigcup T(-\mathbf{u}) m_z L_1,
\end{equation}
where we used the commutator $[T(\mathbf{u}),m_z]=0$ as $\mathbf{u}$ is in $x-y$ plane.
This $L_2'$ can be identified as $L_2(-\mathbf{u})$. Therefore, the ground state energy function has a symmetry $G_0(\mathbf{u}) = G_0 (-\mathbf{u})$, i.e., an in-plane $C_{2z}$ symmetry. 

\paragraph{$m_{1\bar{1}0}$ symmetry of the bilayer.}
If we apply a global $m_{1\bar{1}0}$ operation to the bilayer $L_2$, then the resulting structure is
\begin{equation}
    m_{1\bar{1}0}L_2(\mathbf{u}) = T(m_{1\bar{1}0} \mathbf{u}) m_z L_1 \bigcup L_1 = L_2(m_{1\bar{1}0} \mathbf{u}),
    \label{eq:structure-m1bar10}
\end{equation}
which can be identified as $L_2(m_{1\bar{1}0} \mathbf{u})$. In the derivation of Eq.~\eqref{eq:structure-m1bar10}, we used an internal symmetry of a single layer $m_{1\bar{1}0}L_1 = L_1$, and the commutation relation $m_{1\bar{1}0}T(\mathbf{u}) = T(m_{1\bar{1}0}\mathbf{u})m_{1\bar{1}0}$ for in-plane $\mathbf{u}$.
This proves a $m_{1\bar{1}0}$ symmetry of the ground state energy function $G_0(u) = G_0(m_{1\bar{1}0}\mathbf{u})$.

\paragraph{$m_{110}$ symmetry.} 
$m_{110}$ symmetry directly follows from the two symmetries ($C_{2z}$ and $m_{1\bar{1}0}$) we proved above, due to the multiplicative relation $m_{110}=C_{2z} m_{1\bar{1}0}$.

One should note that the symmetry discussed here is distinct from the space group symmetry in a specific crystal structure, but rather the connection between different stacking structures. In \mzcri, for a generic sliding parameter $\mathbf{u}$, the structure itself can have space group symmetry as low as SG1.

\section{Stacking symmetries in \mote}
We use the analytical tool developed in the previous section to derive symmetries in \mztmote.
Following that notation, a bilayer system of sliding parameter $\mathbf{u}$ is written as $L_2(\mathbf{u}) = T(\mathbf{u})L_1 \bigcup m_z L_1$. 

A monolayer \mote conforms to space group $P2_1$/m, with an in-plane mirror symmetry $m_y$ and full inversion symmetry $\mathcal{P}$. 
\begin{align}
    P L_1 = L_1\\
    T\left(0,\frac{1}{2}\right) m_y L_1 = L_1
\end{align}
This leads to symmetries across sliding structures in the bilayer system \mote considered in the main text.
We now separately consider them. 

\paragraph{\texorpdfstring{$C_{2z}$}{C_2z} symmetry}
Two structures that have sliding parameters $\mathbf{u}$ and $-\mathbf{u}(=C_{2z}\mathbf{u})$ are related by an $m_z$ operation. 
\begin{align*}
    & T(-\mathbf{u}) m_z L_2(\mathbf{u}) \\
    =& T(-\mathbf{u})m_z  m_z L_1 \bigcup T(-\mathbf{u})m_z T(\mathbf{u})L_1 \\
    =& T(-\mathbf{u})L_1 \bigcup m_z L_1\\
    =& L_2(-\mathbf{u})
    \numberthis
\end{align*}
So they have opposite out-of-plane ferroelectric polarization $P_z$ while sharing the same energy and excitation energy response function.

\paragraph{\texorpdfstring{$m_y$}{my} mirror symmetry}
Two structures that have sliding parameters $\mathbf{u}$ and $(=m_y\mathbf{u})$ are related by a $T(0,\frac{1}2)m_y$ operation. 
\begin{align*}
    & T(0,\frac{1}2) m_y L_2(\mathbf{u}) \\
    =& T(0,\frac{1}2) m_y T(\mathbf{u}) L_1 \bigcup T(0,\frac{1}2) m_y m_z L_1\\
    =& T(m_y \mathbf{u})L_1\bigcup m_z L_1
    \numberthis
\end{align*}

\section{Symmetries of response function $\tilde{\chi}_\mathrm{ex}^{ab}$}
Recall from main text Eq.~\eqref{eq:final-expr} that our response function for excitation energy $\tilde{\chi}^{ab}_\mathrm{ex}$ is
\begin{equation}
\tilde{\chi}^{ab}_\mathrm{ex}(\omega)
\;=\;
\frac{i\,\tau\,e^2\,V_\mathrm{u.c.}}{\hbar\,\omega^2}
\sum_{n,l}\int_{\mathbf{k}}
\frac{
    f_{ln}\,\omega_{nl}\,
    v_{nl}^a\,v_{ln}^b
}{
    \omega_{nl} \;-\;\omega \;+\;\tfrac{i}{\tau}
},
\label{eq:chi-ab}
\end{equation}

Under time reversal operation $\mathcal{T}$, which consists of $\mathbf{k}\to-\mathbf{k}$ and complex conjugation $\mathcal{C}$,the energy eigenvalues are symmetric as $\omega_n(\mathbf{k})\to \omega_n(-\mathbf{k})$, the velocity matrix elements transform as $v_{nl}^a(\mathbf{k}) \to -\,v_{ln}^a(-\mathbf{k})$ 
and the response function
\begin{equation}
    \mathcal{T}(\tilde{\chi}^{ab}_\mathrm{ex}) = -\tilde{\chi}^{ba,*}_\mathrm{ex}
\end{equation}
This is expected because the free energy response would be invariant if we also time-reverse the applied field $\omega \leftrightarrow -\omega$. From this result, a time-reversal invariant system would have a symmetric response tensor $\tilde{\chi}^{ab}_\mathrm{ex} = \tilde{\chi}^{ba}_\mathrm{ex}$. 

Under spatial inversion ($\mathcal{P}$), all quantities in Eq.~\eqref{eq:12,deltaG} remains invariant except for a k-space inversion $k \to -k$, which still preserves the overall sum. This is clear also from the fact that the free energy is invariant when the polarizations $\mathbf{E}$ are inverted, because it is a quadratic form in $\mathbf{E}$. 

Under complex conjugation, the frequencies $\omega_{nl}$ and Fermi factors $f_{ln}$ remain invariant, while 
$v_{nl}^a \;\to\; v_{ln}^a$,
and this leads to
\begin{equation}
    \tilde{\chi}^{ab}_\mathrm{ex}(\omega)^* = \tilde{\chi}^{ba}_\mathrm{ex}(-\omega)
\end{equation}
This relation will be useful in the next section where we derive the response for linearly and circularly polarized light.

\section{Response under a linear or circularly polarized light} \label{sec:lpl,cpl}
In Eq.~\eqref{eq:final-expr} in the main text, we gave an expression for a generic input light electric field $\mathbf{E}(\omega)$, while in the examples we only used special cases of either a linearly polarized (LPL) or circularly polarized (CPL) monochromatic light. Here, we explicitly list the actual formulas used. 

Since the excitation free energy
\begin{equation}
    \expval{\Delta G} = \sum_\omega \tilde{\chi}^{ab}_\mathrm{ex}(\omega) E_a(\omega) E_b^*(\omega) 
    \label{eq:12,deltaG}
\end{equation}

\paragraph{Linearly Polarized Light.}
For a beam linearly polarized in-plane along \(\mathbf{n} = (\cos\theta, \sin\theta)\), we use
\begin{equation}
    \begin{cases}
        E_x(\omega) = E_x(-\omega) \;=\; \displaystyle \tfrac{E_0}{2}\cos\theta,\\[6pt]
        E_y(\omega) = E_y(-\omega) \;=\; \displaystyle \tfrac{E_0}{2}\sin\theta.
    \end{cases}
\end{equation}
Substituting this into Eq.~\eqref{eq:12,deltaG}, and using $\tilde{\chi}^{ab}_\mathrm{ex}(\omega)^* = \tilde{\chi}^{ab}_\mathrm{ex}(-\omega) $ gives 
\begin{equation}
    \expval{\Delta G}_{\mathrm{LPL}} = 
                        \mathrm{Re}\left( \tilde{\chi}_\mathrm{ex}^{xx}(\omega) \right) \frac{E_0^2}{2} \cos^2 \theta + 
                        \mathrm{Re}\left( \tilde{\chi}_\mathrm{ex}^{xy}(\omega) + \tilde{\chi}_\mathrm{ex}^{yx}(\omega) \right) \frac{E_0^2}{2} \cos \theta \sin\theta+
                        \mathrm{Re}\left( \tilde{\chi}_\mathrm{ex}^{yy}(\omega) \right) \frac{E_0^2}{2} \sin^2 \theta 
\end{equation}

\paragraph{Circularly Polarized Light.}
For a right-handed circularly polarized beam propagating along \(z\), we take
\begin{equation}
    \begin{cases}
        E_x(\omega) = \displaystyle \tfrac{E_0}{2}, 
        & E_x(-\omega) = \displaystyle \tfrac{E_0}{2},\\[6pt]
        E_y(\omega) = i\, \displaystyle \tfrac{E_0}{2}, 
        & E_y(-\omega) = -\,i\, \displaystyle \tfrac{E_0}{2}.
    \end{cases}
\end{equation}
Substituting these into the same \(\Delta G\) formula yields
\begin{equation}
    \Delta G_{\mathrm{CPL}}
    \;=\;
        \mathrm{Re}\left( \tilde{\chi}_\mathrm{ex}^{xx}(\omega) \right) \frac{E_0^2}{2}+
        \mathrm{Im}\left( \tilde{\chi}_\mathrm{ex}^{xy}(\omega) - \tilde{\chi}_\mathrm{ex}^{yx}(\omega) \right) \frac{E_0^2}{2}+
        \mathrm{Re}\left( \tilde{\chi}_\mathrm{ex}^{yy}(\omega) \right) \frac{E_0^2}{2} 
\end{equation}

\section{Effect of other optical response mechanisms}
The contribution introduced in \eqref{eq:final-expr} in the main text is the most dominant among other possible effects that are quadratic in $\mathbf{E}$ for inducing ionic motion and thus structural transition. We compare nonlinear excitation energy with the energy from a few other physical mechanisms.
\paragraph{The field interaction energy.}
In the main text, we derived two terms that can contribute to the free energy change $G(\mathbf{u},\mathbf{E}) - G_0(\mathbf{u})$: Eq.~\eqref{eq:field-energy} and Eq.~\eqref{eq:final-expr}.  Eq.~\eqref{eq:field-energy} the energy produced by the light field interacting with the linear response (what we call the field-interaction energy). Here, we do an order of magnitude estimate for the two terms to show the relative strengths of the two terms.

For both terms, there is a common denominator can be approximated, while near resonance $|\omega-\omega_{nl}| \ll {\tau}^{-1}$, as 
\begin{equation}
    \frac{1}{\omega_{ln} - \omega + \frac{i}{\tau}} \approx \mathrm{P} \left(\frac{1}{\omega_{ln} - \omega}\right) - i\pi \delta(\omega_{ln} - \omega)
\end{equation}
Here, only the imaginary part has a non-vanishing contribution to the final free energy expression. Using also the relation $\sum_{nl} \delta(\omega_{ln} - \omega) = D(\omega)$, 
Eq.~\eqref{eq:field-energy} can be estimated as $\frac{e^2 v^2 V_\mathrm{u.c.} D(\omega) E^2}{\hbar \omega^2}$, and the nonlinear excitation energy Eq.~\eqref{eq:final-expr} is larger by a factor of $\omega\tau \gg 1$. 

\begin{figure}
    \centering
    \includegraphics[width=0.5\linewidth]{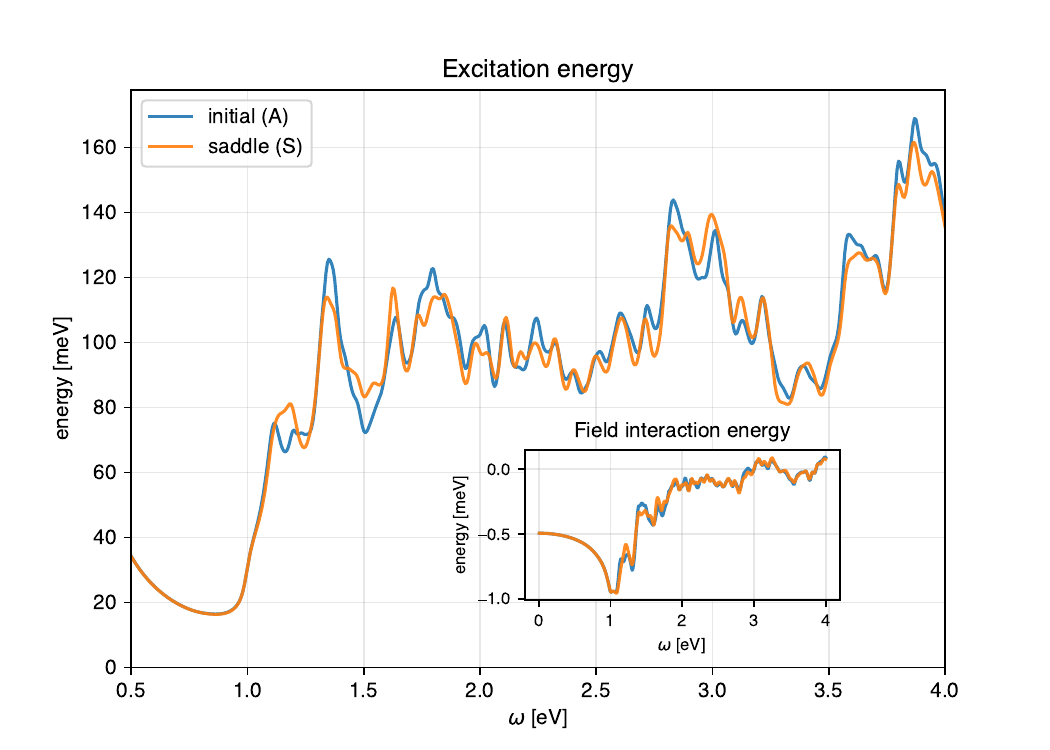}
    \caption{Comparison of nonlinear excitation Eq.~\eqref{eq:interaction-energy} and the field energy Eq.~\eqref{eq:field-energy} terms at stackings $A$ and $S$.}
    \label{fig:ex_energy_AS}
\end{figure}

To validate this numerically, we compute both terms from our DFT data for \mzcri. 
Specifically, we calculate the optical frequency dependence of both terms in the free energy for stacking A and S, shown in Fig. \ref{fig:ex_energy_AS}.
For both stackings, the nonlinear excitation energy calculated in the main text is $\sim10^2$ larger than the field interaction energy, as expected from the order of magnitude estimation above.

\begin{figure}
    \centering
    \includegraphics[width=0.5\linewidth]{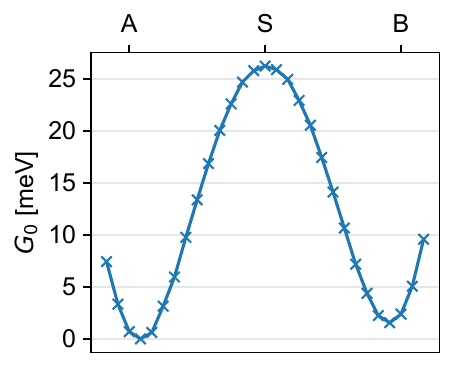}
    \caption{Ground state energy $G_0(\mathbf{u})$ of \mzcri for stackings along A-S-B path. $A$, $S$, $B$ are the three stackings labeled in Fig.~\ref{fig:1} in Main Text.}
    \label{fig:mz_cri3_pathLandscape}
\end{figure}

\paragraph{The validity of neglecting direct phonon resonance: phonon frequency in \cri.}
From the ground state energy curve along transition path (shown in Fig.~\ref{fig:mz_cri3_pathLandscape}), we can estimate the frequency of ionic motion and therefore the importance of direct ion-light coupling. 

As an example, we calculate the interlayer vibrational frequency along the transition path for stacking A. Since it is the softest degree of freedom, it is also the minimum optical phonon frequency in this system. 
From Fig.~\ref{fig:mz_cri3_pathLandscape}, we can quadratically fit $G_0(\mathbf{u})$ near stacking $A$ to get $\frac{\partial^2 G_0(u)}{\partial u^2} = 204 \, \mathrm{meV/Å}$. The effective mass for this vibration is $M = 864.5u$. Hence the frequency is 
\begin{equation}
    \omega_{0,\mathrm{ion}} = \sqrt{\frac{\frac{\partial^2 G_0(u)}{\partial u^2}}{M}} = 2\pi \times 0.14 \, \mathrm{THz}
\end{equation}
which is $\sim 10^3$ off-resonant with light used in this work $\omega \sim 2\pi \times 700 \,\mathrm{THz}$. The interlayer sliding induced directly by light is then $x \sim \frac{ZeE_0 }{m} \frac{1}{|\omega_0^2 - \omega^2|} \sim 10^{-18}\mathrm{m}$. Since a structural transition requires layer motion of $\sim 10^{-9}\,\mathrm{m}$, this far off-resonant effect can be safely neglected. 

\section{Excited state band structure in \mzcri}
\begin{figure}[ht]
    \centering
    \includegraphics[width=0.8\linewidth, page=5]{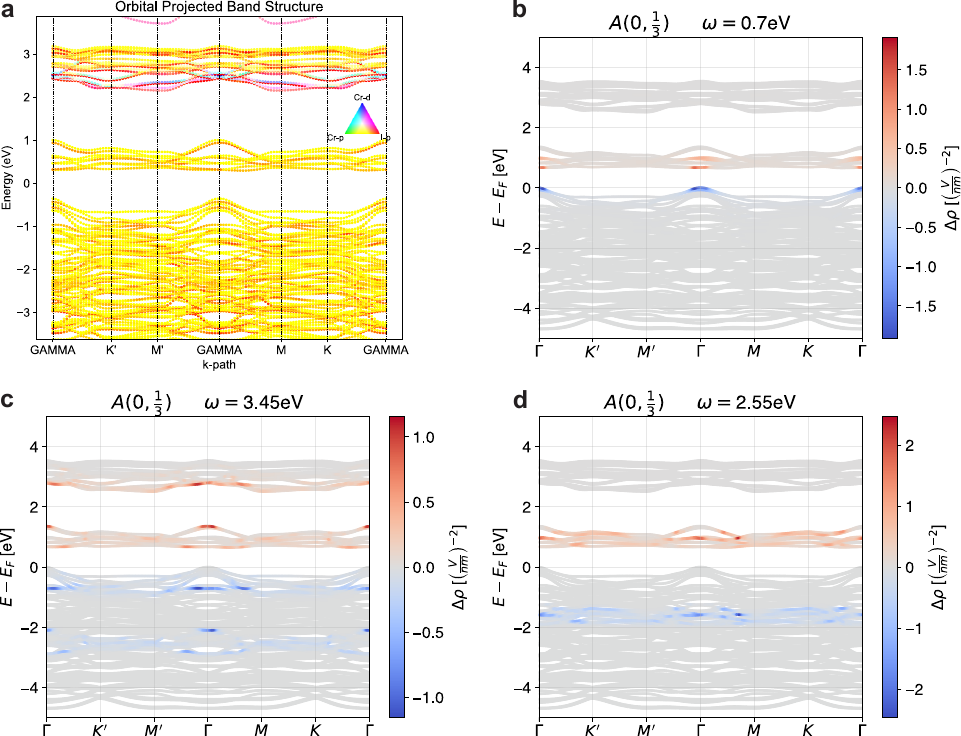}
    \caption{Excited state and orbital components of \mzcri. (a) A bandstructure plot of\mzcri showing orbital constitution by color: Cr-d orbital in blue, Cr-p orbital in green, and I-p orbital in red, hybrid orbitals are mapped to color by additive color mixing. (b)(c)(d) bandstructure plot showing electron excitation under 0.7eV, 3.45eV, and 2.55eV light, from regions (in blue) of the valence band to regions (in red) of the conduction band. The k-point labels $\Gamma$, $M$, $K$ follow conventional definitions of high-symmetry points in a hexagonal Brillouin zone. $M'$ and $K'$ are time-reversed ($\mathbf{k} \to -\mathbf{k}$) counterparts of $M$ and $K$.}
    \label{fig:excited_bandstructure}
\end{figure}

Here we explicitly evaluate the excited state bandstructure to visualize these excitations. In Fig.~\ref{fig:excited_bandstructure} (b)(c)(d), we show the excited state population for one frequency just above bandgap ($\hbar\omega=$0.7eV), and two frequencies promoting $A\to B$ and $B\to A$ transitions considered in the main text. The covalent nature of the relevant bands (Fig.~\ref{fig:excited_bandstructure} (a)) are conducive to a large quantum metric, and thus a large NELIC effect. 

\section{Effect of electronic relaxation time $\tau$} \label{sec:effect-of-tau}
\begin{figure}
    \centering
    \includegraphics[width=0.5\linewidth]{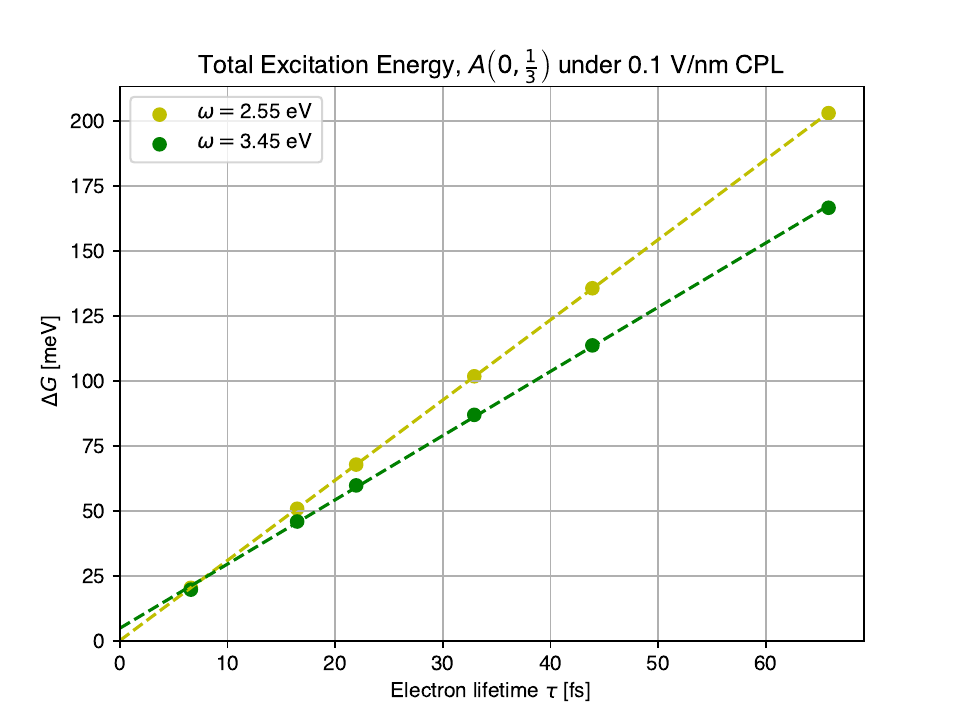}
    \caption{NELIC energy calculated for different choices of $\tau$. Dots are data points calculated from individual DFT data. Broken lines are linear fits.}
    \label{fig:envsTau}
\end{figure}

In the calculations done in the main text, we set the electronic relaxation time as a universal empirical constant $\tau = \frac{\hbar}{30 \mathrm{meV}}$. Here, we evaluate other possibilities of this value and its effect on the excitation energy, shown in Fig. \ref{fig:envsTau}.

From Fig.~\ref{fig:envsTau}, we conclude that in the regime considered in this work, NELIC energy is to a very well approximation proportional to $\tau$ in main text. This can be understood by taking the large-$\tau$ (clean) limit in Eq.~\eqref{eq:final-expr} in the main text. 

\section{Effect of using short pulses}
Practically in an experiment, short pulses are more often used instead of continuous waves to avoid heating-related damages. Here, we demonstrate how Eq.~\eqref{eq:final-expr} in the main text, originally derived by assuming continuous wave excitation, can be extended to the femtosecond pulse regime. A consequence is that under a femtosecond pulse excitation, we can still trigger ferroic order switching in \mzcri.

When excited by a pulse, the electronic excitation number accumulated over $t_\mathrm{pulse}$ follow Fermi's golden rule
\begin{equation}
    \rho_n  = \frac{2\pi t_\mathrm{pulse}}{\hbar^2} \sum_l |V_{ln} (\omega)|^2 \delta(\omega_{nl} - \omega)
    \label{eq:fermi-golden}
\end{equation}
where $l$ ($n$) is a valence (conduction) band index. In Sec.~\ref{sec:effect-of-tau}, we argued that in the regime $\tau$ can be considered large, or that the level broadening is sharp. We can then replace the delta function in \eqref{eq:fermi-golden} as a Lorentzian with width $\tau$ to get
\begin{equation}
    \rho_{n} = -\frac{i t_{\mathrm{pulse}}}{\hbar^2} \sum_l \left( \frac{f_{nl} |V_{nl}(\omega)|^2}{\omega_{ln} - \omega + i/\tau} - \frac{f_{ln} |V_{ln}(\omega)|^2}{\omega_{nl} - \omega + i/\tau} \right)
\end{equation}
which is exactly Eq.~\eqref{eq:final-expr} in the main text multiplied by a factor $\frac{t_\mathrm{pulse}}{\tau}$. 
Therefore, a pulse of duration $t_\mathrm{pulse} = \tau = 22\,\text{fs}$ can already pump electrons to the same level of excitation as a continuous wave with the same intensity and polarization does. Past the pulse duration, electrons will remain in the same excited state for $T_\mathrm{decay}$ before relaxing to the ground state through radiative decay. During this time, the ions can experience the same NELIC force, so that the ferroic order switching can happen. Here, we calculate and compare the characteristic time for ions to switch ferroic state $T_\mathrm{switch}$ and the radiative decay time $T_\mathrm{decay}$. 

\begin{figure}
    \centering
    \includegraphics[width=0.75\linewidth]{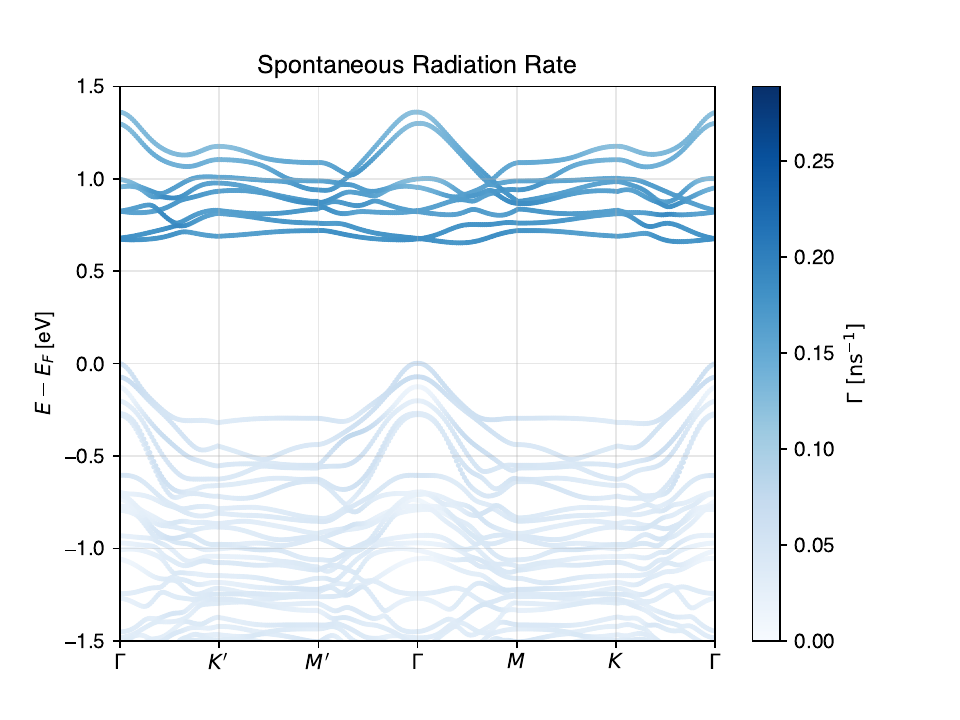}
    \caption{Spontaneous decay rate for Bloch states of \mzcri at stacking A, represented by colors of a bandstructure plot.}
    \label{fig:sponRate}
\end{figure}

\paragraph{Radiative decay time. }
This decay rate can be calculated with
\begin{equation}
 \Gamma_m = \frac{4\alpha}{3c^2} \sum_{n < m} \omega_{mn} |v_{mn}|^2 \sqrt{\text{Re}\, \varepsilon_r(\omega_{mn})} 
 \label{eq:sponRate}
\end{equation}
where $\Gamma_m$ is the decay rate of state $m$, $\alpha\approx 1/137$ is the fine-structure constant, $c$ is the speed of light in vacuum, $\varepsilon_r (\omega_{mn})$ is the relative dielectric function at frequency $\omega_{mn}$. By an order of magnitude estimation taking $\omega_{mn} = 1 \,\mathrm{eV/}\hbar$, $v_{mn} = \frac{1\mathrm{nm}}{\omega_{mn}}$, this spontaneous rate can be found to be $\Gamma \sim 1 \mathrm{ns}^{-1}$.
The same conclusion can be reached by evaluating \eqref{eq:sponRate} for Bloch states along a high-symmetry path in Brillouin Zone of \mzcri at stacking A, as shown in Fig. \ref{fig:sponRate}.
Therefore, the radiative decay time $T_\mathrm{decay} \sim 1 \mathrm{ns}$.

\paragraph{Structural transition time. }
With the electrons being in the excited state, the system moves on a stacking-dependent potential. The time for the system to move from the initial (A) stacking to the saddle point (S) can be estimated with \cite{landau_mechanics_2013}
\begin{equation}
T_{\text{ion}} = \int_\mathrm{A}^\mathrm{S} \frac{\dd x}{v(x)} = \int_\mathrm{A}^\mathrm{S} \sqrt{\frac{M}{2 \left(\Phi(x_\mathrm{A}) - \Phi(x)\right)}} \, \dd x 
\label{eq:t-ion}
\end{equation}
After this time, the structure shifts beyond the saddle point so the ionic potential does not need to remain in the excited shape: the ionic potential at electronic ground state suffices to drive the remaining ionic motion. So this is the minimum time required for electrons to stay excited. 

From \eqref{eq:t-ion}, $T_\mathrm{ion}$ evaluates to 1.40 ps for $\hbar\omega = 3.45\,\mathrm{eV}$ at $1.5$ times the critical $\mathrm{A}\to\mathrm{B}$ transition intensity, and $1.82 \,\mathrm{eV}$ for $\hbar\omega = 2.55 \, \mathrm{eV}$ at the $1.5$ times the $\mathrm{B}\to\mathrm{A}$ critical intensity. This $T_\mathrm{ion}\sim\mathrm{ps}$ order of magnitude is close to one cycle of phonon oscillation as expected. 

Because the structural transition takes $\sim\mathrm{ps}$ to happen, which is much shorter than electronic excitation decay time $T_{\mathrm{decay}}$, we expect that the initial $t_\mathrm{pulse}=22\,\mathrm{fs}$ pulse can trigger a ferroic switching despite the short pulse duration.

\section{Computational Details}
All terms in Eq.~$\eqref{eq:final-expr}$, except $\gamma_{nl}=\gamma$ which is taken to be uniformly 30 $\frac{\mathrm{meV}}{\hbar}$, are evaluated using \textit{ab initio} methods in the two bilayer 2D materials. We utilized the Vienna ab initio simulation package (VASP) to perform first-principles calculations based on density functional theory (DFT). To handle exchange-correlation interactions, we applied the generalized gradient approximation (GGA) in the form of Perdew-Burke-Ernzerhof (PBE). The core and valence electrons were respectively treated with the projector augmented wave (PAW) method and plane-wave basis functions, with an energy cutoff of 400eV. For DFT calculations, a $\Gamma$-centered $k$ mesh was used to sample the first Brillouin zone, with a grid of  7×7×1. To treat the $d$ orbitals of spin-polarized Cr atoms in \cri, we employed the DFT+U method, setting $U$ to 3.0 eV. We used the Wannier90 package to generate tight-binding orbitals by projecting Bloch waves in DFT calculations onto Cr-d, I-p orbitals for \cri, and Mo-d, Te-p orbitals for \mote. The tight-binding Hamiltonian was then used to interpolate the band structure on a 320×320×1 $k$ mesh to calculate the dielectric function, band energy, and density response functions. The $k$-mesh convergence is well-tested. Non-collinear spin-orbit coupling is included in all calculations.
We obtained the bilayer \cri system by adding a 10Å vacuum layer on top of a bulk bilayer unit cell. For each stacking $\mathbf{u}=(u_x,u_y)$, the in-plane coordinates were fixed while the interlayer distance was allowed to relax before subsequent calculations are performed. Relaxation was carried out until the force on each atom $|F_i| < 10^{-3} \textrm{eV/Å}$. For readability, we smoothed over the free energy values in Fig.~\ref{fig:2}(e) and Fig.~\ref{fig:2}(f) with Gaussian-smoothing with radius of 2 nearest-neighbor stackings. 

\end{document}